\documentclass[aps,prb,twocolumn]{revtex4-1}
\usepackage{amssymb}
\usepackage{amsmath}
\usepackage{graphicx}
\usepackage{epsfig}
\usepackage{color}

\begin{document}

\title{Localization transitions and mobility edges in quasiperiodic ladder}
\author{R. Wang}
\author{X. M. Yang}
\author{Z. Song}
\email{songtc@nankai.edu.cn}
\affiliation{School of Physics, Nankai University, Tianjin 300071, China}

\begin{abstract}
We investigate localization properties in a two-coupled uniform chains with
quasiperiodic modulation on interchain coupling strength. We demonstrate
that this ladder is equivalent to a Aubry-Andre (AA) chain when two legs are
symmetric. Analytical and numerical results indicate the appearance of
mobility edges for asymmetric ladder. We also propose an easily engineered
quasiperiodic ladder system which is a moir\'{e} superlattice system
consisting of two-coupled uniform chains. An irrational lattice constant
difference results in quasiperiodic structure. Numerical simulations show
that such a system supports mobility edge. Additionally, we find that the
mobility edge can be detected by a dynamic method, which bases on the
measurement of surviving probability in the presence of a single imaginary
negative potential as a leakage. The result provides insightful information
about the localization transitions and mobility edge in experiment.
\end{abstract}

\maketitle

\section{Introduction}

\label{sec_intro}

The localization phase in quantum systems, which is originally rooted in
condensed matter \cite{PA}, has recently attracted a lot of theoretical and
experimental interest in a variety of fields, including light waves in
optical random media \cite{ST,SG,CC,DM}, matter waves in optical potentials
\cite{KD,GR,FJ,GS}, sound waves in elastic media \cite{HH}, and quantum
chaotic systems\cite{JC}, since the localization of quantum particles could
prevent the transport necessary for equilibration in isolated systems.
Anderson localization\cite{PW} predicts that single-particle wave functions
become localized in the presence of some uncorrelated disorder, leading to a
metal-insulator transition caused by the quantum interference in the
scattering processes of a particle with random impurities and defects. A
conventional Anderson localization is not controllable in one and
two-dimensional systems. Nevertheless, localization does not require
disorder and fortunately, the Aubry-Andre model \cite{MY,SA}, which has
quasiperiodic potential, exhibits a transition between a localized and
extended phases. In practice, quasiperiodic potentials arise naturally in
optical experiments using lasers with incommensurate wave vectors.
Accordingly, many experiments in such systems have now observed
single-particle localization \cite{LDN,LF,GR,YL,GM,MS}. The localization
transition is always an attractive topic, especially in one-dimensional or
quasi-one-dimensional system.

While most studies in this field have focused on the model with random or
quasiperiodic on-site potential, the localization arising from the
modulation of tunneling is far from being well understood and we believe
that the essence of localization occurring in nature involve matters with
random or quasiperiodic structure. Recently there has been a growing
interest in the influence of the moir\'{e} pattern in physical systems. The
moir\'{e} pattern as a new way to apply periodic potentials in van der Waals
heterostructures to tune electronic properties, has been extensively studied
\cite{Ponomarenk,Dean,Hunt,Gorbachev,Song,Jung}. Many interesting phenomena
have been observed in the heterostructure\ materials with small twist angles
and mismatched lattice constants. Moir\'{e} patterns in condensed matter
systems are produced by the difference in lattice constants or orientation
of two two-dimensional lattices when they are stacked into a two-layer
structure. In principle, a moir\'{e} system can be deliberately designed
(engineered) as\ a quasiperiodic structure when the difference of two
ingredients is incommensurate. It is interesting to study the Anderson
localization in a moir\'{e}-like system, which provides another way to
demonstrate edge mobility\ arising from the quasiperiodic ladder structure.
An advantage of this scheme is the simplicity of the model which should be
feasible to engineered in practice. The aim of this paper is to construct an
easily engineered quasiperiodic\ system which possessing mobility edges.

In the paper, we study a quasiperiodic ladder system, which consists of two
uniform chains. The quasiperiodicity arises from the quasiperiodic
modulation on interchain coupling strength. At first, we consider the case
with a quasiperiodic\ modulation on the rung of the ladder. We show that
this ladder is equivalent to a Aubry-Andre (\textrm{AA}) chain when two legs
are symmetric. Analytical and numerical results indicate the appearance of
mobility edges for asymmetric ladder. Secondly, we propose an easily
engineered quasiperiodic ladder system. The quasiperiodic structure arises
from the slight difference of lattice constants of two legs, rather than
deliberately engineered. It is a moir\'{e}-superlattice-like system with
small irrational\ difference between the top and bottom chain. Numerical
simulations show that localization transitions and mobility edges can be
observed in such a quasiperiodic ladder system. Finally, we employ the
fidelity of quantum states and the measurement of surviving probability\ to
detect localization transition and mobility edge through a dynamic way. The
model setup has relative simple geometrical structure and contains zero
potentials. It should be experimentally accessible in atomic optical
lattices and photonic waveguides.

This paper is organized as follows. In Sec. \ref{Aubry-Andre ladder}, we
present the model Hamiltonian and analyze the structure of the lattice. In
Sec. \ref{Moire ladder}, we propose an easily engineered quasiperiodic
ladder system. Sec. \ref{dynamical detection of mobility edge} devotes to
the numerical simulation of the model, revealing the dynamical detection of
mobility edge.\ Finally, we give a summary and discussion in Sec. \ref%
{summary}.

\section{Aubry-Andre ladder}

\label{Aubry-Andre ladder}

The quasiperiodic structure in a standard \textrm{AA} chain arises from the
quasiperiodic on-site potential. A nature conjecture is that the appearance
of localization is introduced by the quasiperiodic structure rather than the
on-site potential only. In this section we demonstrate this point by
considering a AA ladder system, in which quasiperiodic structure arises from
the rung hopping strength. We will show that such a system can have
localized eigenstates.

\begin{figure}[tbp]
\centering
\includegraphics[bb=127 237 421 560,width=0.4\textwidth, clip]{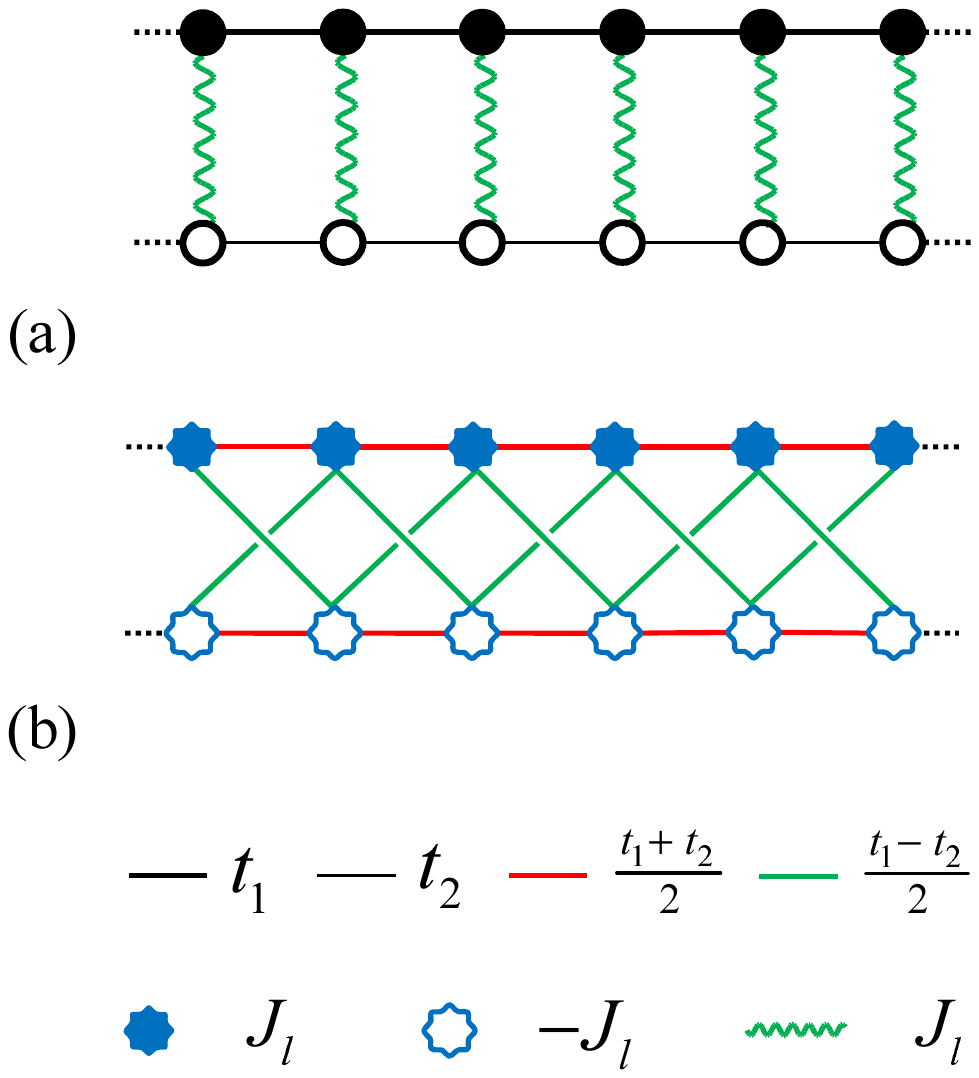}
\caption{(Color online) Schematic illustration of the transformation from a
ladder with quasiperiodic hopping to that with on-site potential. (a) A
ladder system with quasiperiodic interchain hopping, which is indicated by a
wavy line. (b) A ladder system with quasiperiodic on-site potential, which
is indicated by a wavy circles, but uniform hopping strengths. Two legs are
coupled uniformly by interleg in a crossing manner.}
\label{fig1}
\end{figure}

The Hamiltonian is composed by three parts,
\begin{equation}
H=H_{1}+H_{2}+H_{\mathrm{R}}\mathrm{,}  \label{H}
\end{equation}%
where the Hamiltonian of legs is
\begin{equation}
H_{\lambda }=\sum\limits_{l}t_{\lambda }\left\vert l,\lambda \right\rangle
\left\langle l+1,\lambda \right\vert +\mathrm{H.c.,}
\end{equation}%
with $(\lambda =1,2),$ and the rung term is
\begin{equation}
H_{\mathrm{R}}=\sum_{l=1}^{N}J_{l}\left\vert l,1\right\rangle \left\langle
l,2\right\vert +\mathrm{H.c.},
\end{equation}%
with the rung hopping strength%
\begin{equation}
J_{l}=J_{\mathrm{R}}\cos (2\pi lb).
\end{equation}%
When $b$ takes irrational number, for example $b=2/(\sqrt{5}-1)$, $J_{l}$\
is a quasiperiodic function. Notably, it is equivalent to a model with
quasiperiodic on-site potential.\ Thus, we rewrite $H_{\mathrm{R}}$ as the
form
\begin{equation}
H_{\mathrm{R}}=\sum_{l=1}^{N}\left( J_{l}\left\vert l,+\right\rangle
\left\langle l,+\right\vert -J_{l}\left\vert l,-\right\rangle \left\langle
l,-\right\vert \right) ,
\end{equation}%
where the bond and anti-bond states are
\begin{equation}
\left\vert l,\pm \right\rangle =\frac{1}{\sqrt{2}}\left( \left\vert
l,1\right\rangle \pm \left\vert l,2\right\rangle \right) .
\end{equation}%
Then we have
\begin{equation}
H=H_{+}+H_{-}+H_{\mathrm{C}},  \label{H2}
\end{equation}%
where
\begin{equation}
H_{\pm }=\sum_{l=1}^{N}[\frac{t_{1}+t_{2}}{2}\left( \left\vert l,\pm
\right\rangle \left\langle l+1,\pm \right\vert +\mathrm{H.c.}\right) \pm
J_{l}\left\vert l,\pm \right\rangle \left\langle l,\pm \right\vert ]
\end{equation}%
describes two independent chains with opposite on-site potential $\pm J_{l}$%
, $H_{\pm }$\ obey the relation $\left[ H_{+},H_{-}\right] =0$, and%
\begin{equation}
H_{\mathrm{C}}=\frac{t_{1}-t_{2}}{2}\sum_{l=1,\sigma =\pm }^{N}(\left\vert
l,\sigma \right\rangle \left\langle l+1,-\sigma \right\vert )+\mathrm{H.c.}
\end{equation}%
describes interchain crossing hopping term. We would like to point out that
the above transformation is independent of the form of $J_{l}$, periodic or
not. In\ Fig. \ref{fig1}. we sketch the lattice geometries of the
Hamiltonian from Eqs. (\ref{H}) and (\ref{H2}).
\begin{figure}[tbph]
\centering
\includegraphics[width=0.5\textwidth, clip]{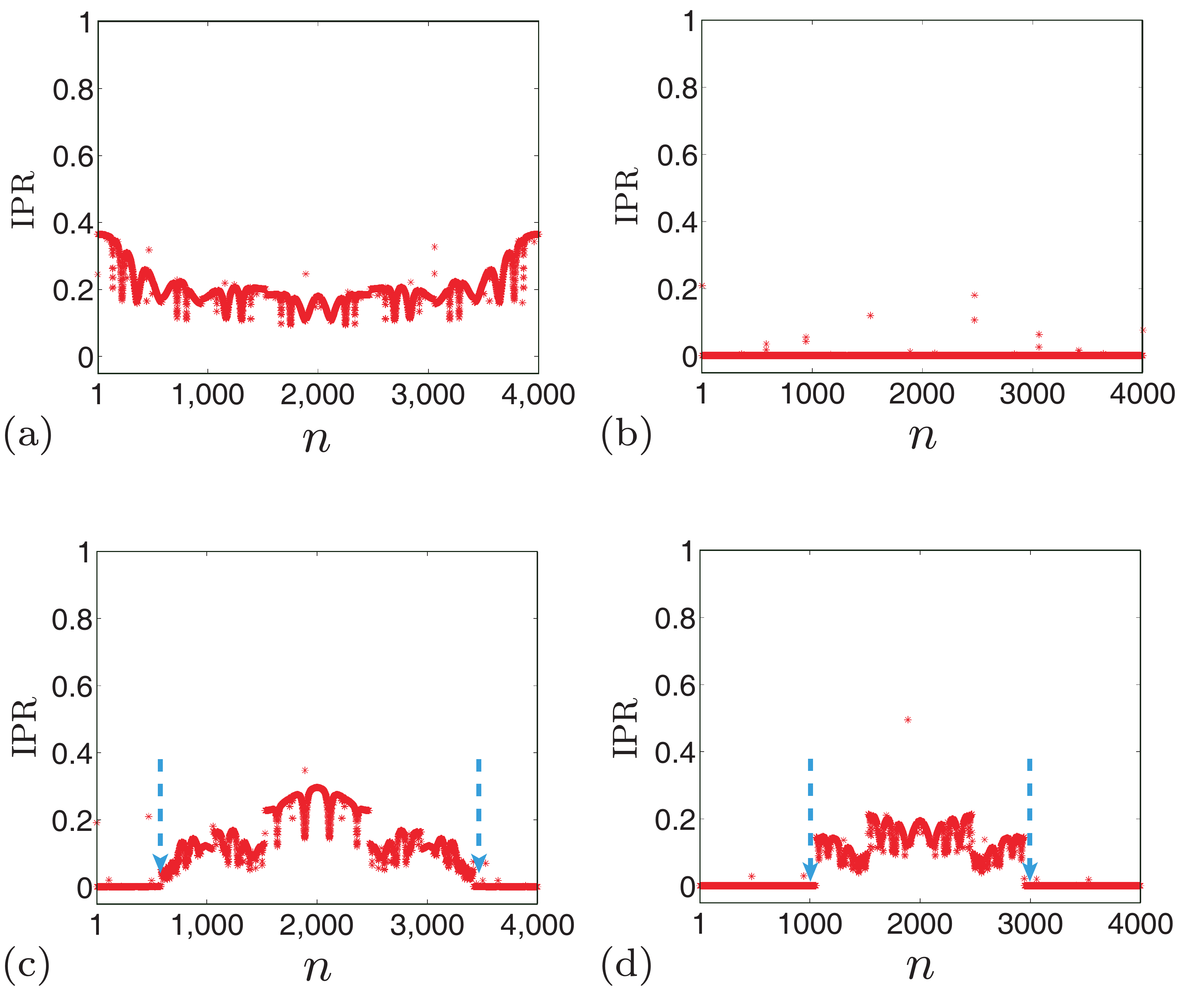}
\caption{(Color online) Numerical simulations for the ladder system from Eq.
(\protect\ref{H}). Plots of IPRs along energy levels for four typical sets
of parameters: (a) $t_{1}=t_{2}=0.4J_{\mathrm{R}}$; (b) $t_{1}=t_{2}=0.8J_{%
\mathrm{R}}$;\ (c) $t_{1}=0.2J_{\mathrm{R}}$, $t_{2}=0.4J_{\mathrm{R}}$. (d)
$t_{1}=0.2J_{\mathrm{R}}$, $t_{2}=0.8J_{\mathrm{R}}$.\ It shows that (a)
indicate all eigenstates are localized, while (b) indicate all eigenstates
are extended. Panel (c) and (d) shows different patterns, which contains two
different regions with small and large IPRs, respectively. The blue arrows
indicate positions of mobility edges, which separate into two different
regions. The size of system is $N=2000$, parameter is $J_{\mathrm{R}}=5$\
and $n$ stands for the $n$th of energy level.}
\label{fig2}
\end{figure}
We concentrate on two cases. (i) For the case with $t_{1}=t_{2}$, we have $%
H_{\mathrm{C}}=0$, i.e., the ladder system is equivalent to two independent
\textrm{AA} chains. There is no mobility edge, and the localization
transition occurs at $t_{1}=t_{2}=0.5J_{\mathrm{R}}$. (ii) For the case with
$t_{1}\neq t_{2}$, $H$\ can be regarded as two coupled \textrm{AA} chains.
Previous works \cite%
{DSM,DSM2,Thouless,DSM3,J.B.,J.B.2,Sr.Ga.,MJ,M.L.,MJ2,A.P.,L.G.,S.G.}
indicate that it is possible to construct models with quasiperiodic
potentials, which do exhibit exact mobility edge even in one dimension.
Thus, it is expected that the mobility edge may appear for an asymmetric
ladder. We perform numerical simulation to explore mobility edge within a
range of parameters. To characterize the localization features, we employ
the inverse participation ratio (\textrm{IPR}) of eigenstate $\left\vert
\psi _{n}\right\rangle $, as a criterion to distinguish the extended states
from the localized ones, which is defined as\textbf{\ }%
\begin{equation}
\mathrm{IPR}^{(n)}=\frac{\sum_{l}\left\vert \left\langle \psi
_{n}\right\vert l\rangle \right\vert ^{4}}{(\sum_{l}\left\vert \left\langle
\psi _{n}\right\vert l\rangle \right\vert ^{2})^{2}}.
\end{equation}%
For spatially extended states, it approaches to zero, whereas it is finite
for localized states. We plot the results in Fig. \ref{fig2}, which
indicates the mobility edge appears when suitable parameters are taken. We
conclude that it is possible to construct a ladder system with quasiperiodic
interchain hopping strength, which also exhibits mobility edge.
\begin{figure}[tbp]
\centering
\includegraphics[bb=23 115 563 697,width=0.45\textwidth, clip]{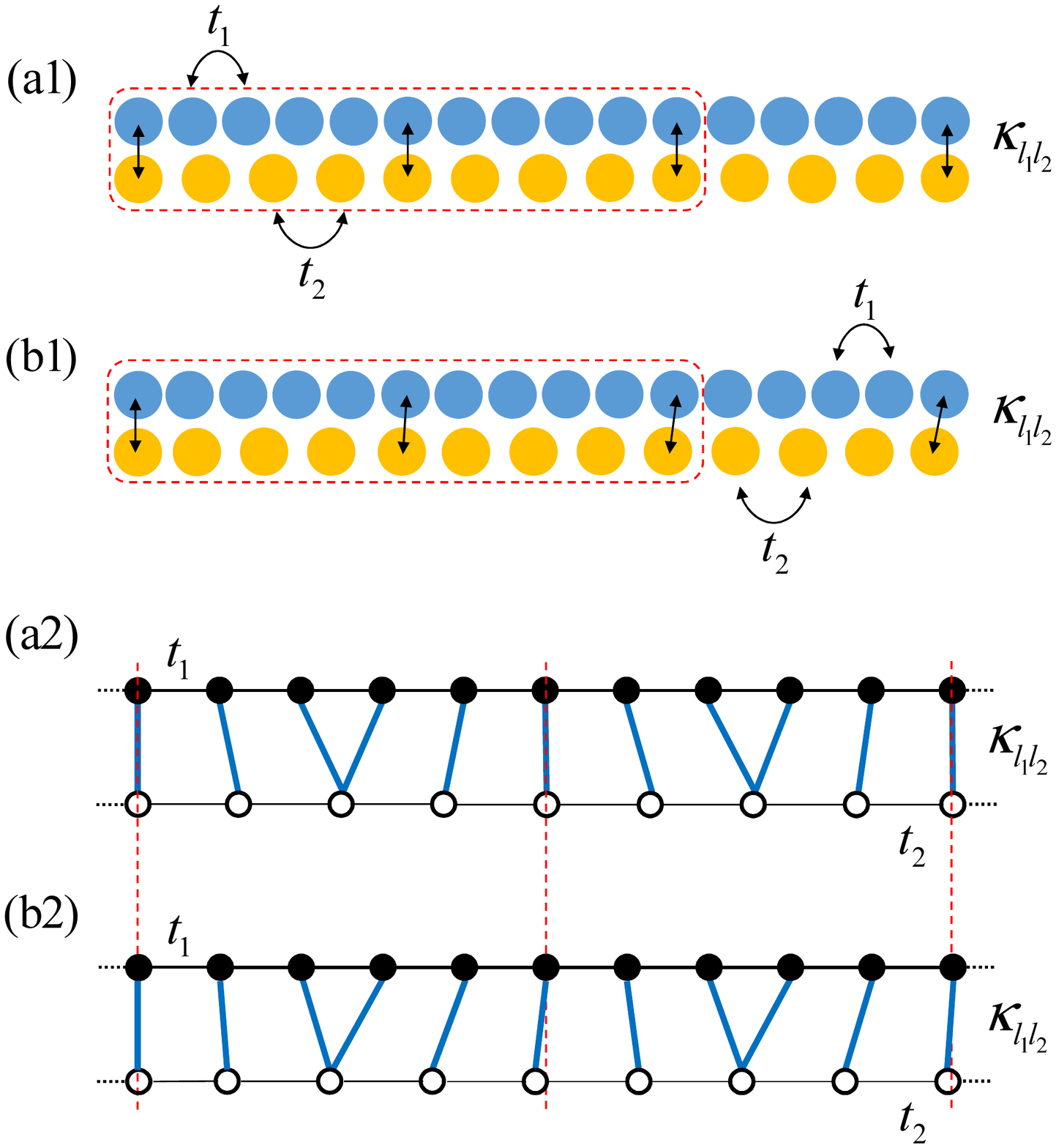}
\caption{(Color online) Schematic illustration of the moir\'{e} two-leg
ladder system. It consists of two coupled uniform chains. The irregular
structure arises from the difference of lattice constants ($1$ and $1+\Delta
$) between two legs. There are two types of ladder structures, which have a
slight difference: (a1) \ a rational number, for example $\Delta =0.25$;
(b1) a irrational number, for example $\Delta =(\protect\sqrt{5}-1)/5\approx
0.25$. The interleg hopping strength across $l_{1}$th site in leg $1$ and $%
l_{2}$th site in leg $2$ is indicated by $\protect\kappa _{l_{1}l_{2}}$,
which is distance dependent (see Eq. (\protect\ref{Kappa 12})). (a2) and
(b2) sketch the main hopping strengths for the structures with two
(quasi)periods, as the schematics of the parts in (a1) and (b1) marked by
dashed red rectangles. Here vertical dashed lines are added in each panel to
indicate the difference.}
\label{fig3}
\end{figure}
\begin{figure}[tbp]
\centering
\includegraphics[width=0.5\textwidth, clip]{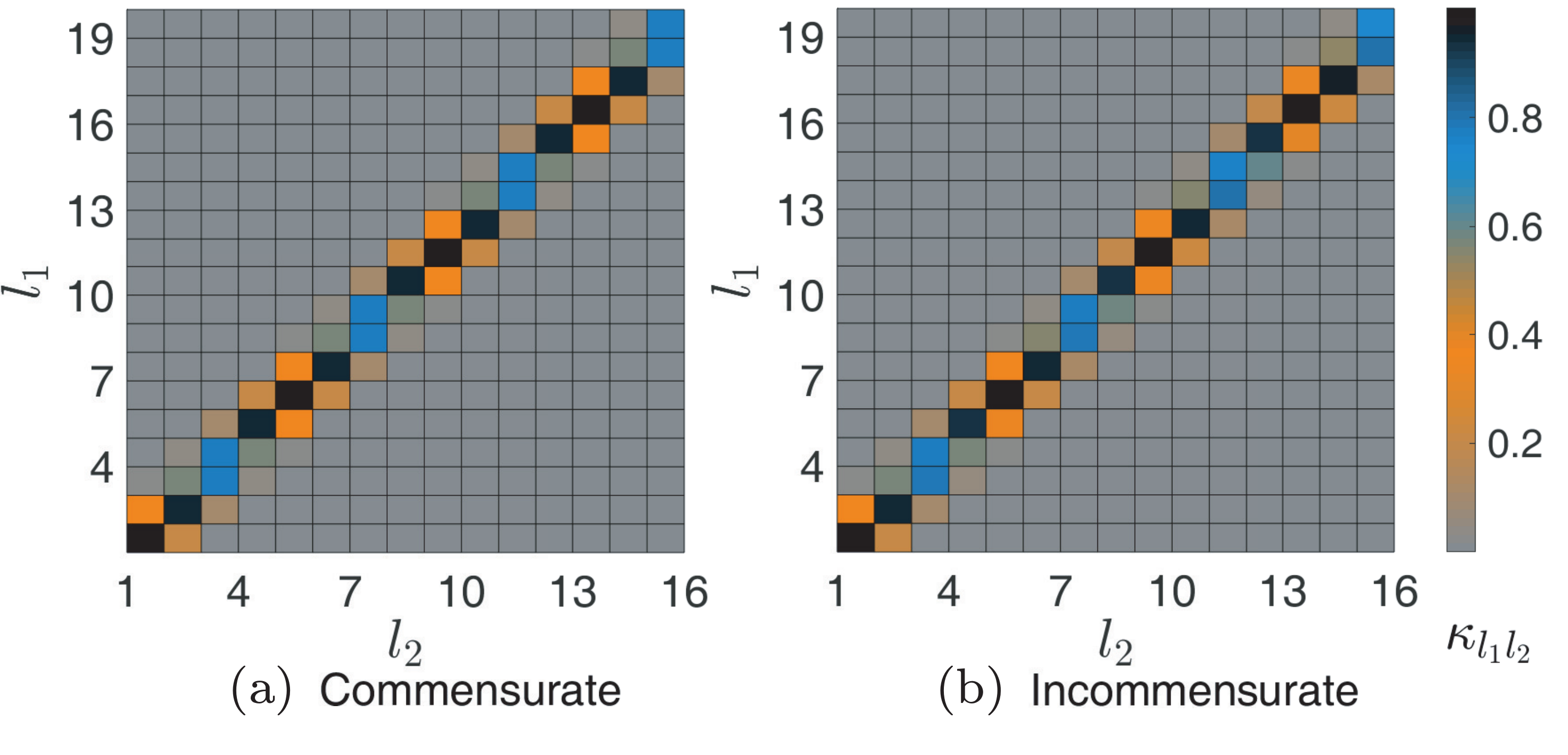}
\caption{(Color online) Color contour maps for the profiles of $\protect%
\kappa _{l_{1}l_{2}}$\ are obtained from Eq. (\protect\ref{Kappa 12}) with
(a) $\Delta =0.25$, (b) $\Delta =(\protect\sqrt{5}-1)/5$. Plots contain
three (quasi)periods. Two patterns have a slight difference: (a) is exact
periodic while (b) is quasiperiodic. The parameters are $\protect\alpha =3$,
$t_{1}=t_{2}=1$, and $\protect\kappa _{0}=1$.}
\label{fig4}
\end{figure}
\begin{figure*}[tbph]
\begin{minipage}{0.32\linewidth}
\centerline{\includegraphics[width=5.5cm]{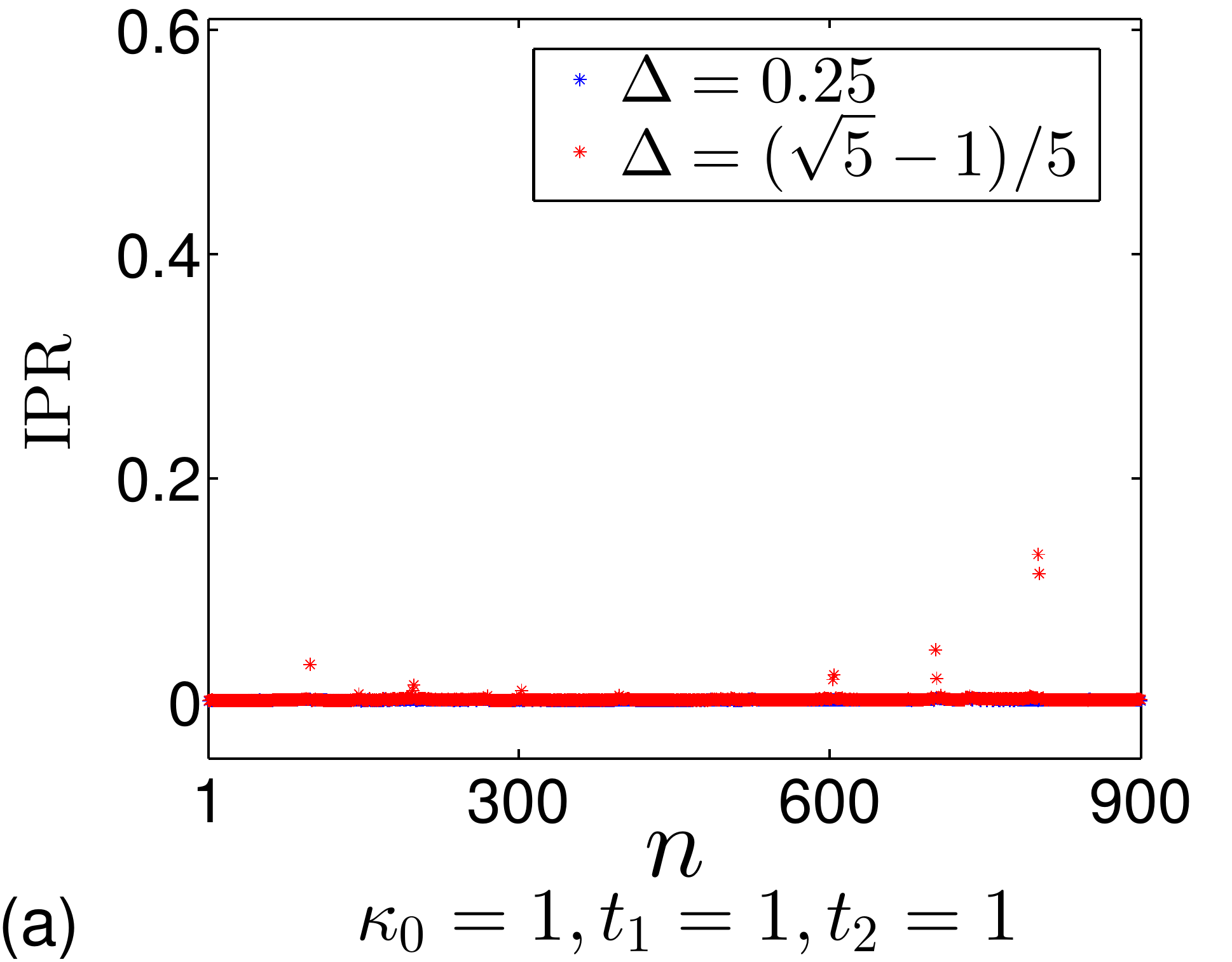}}
\end{minipage}
\begin{minipage}{0.32\linewidth}
\centerline{\includegraphics[width=5.5cm]{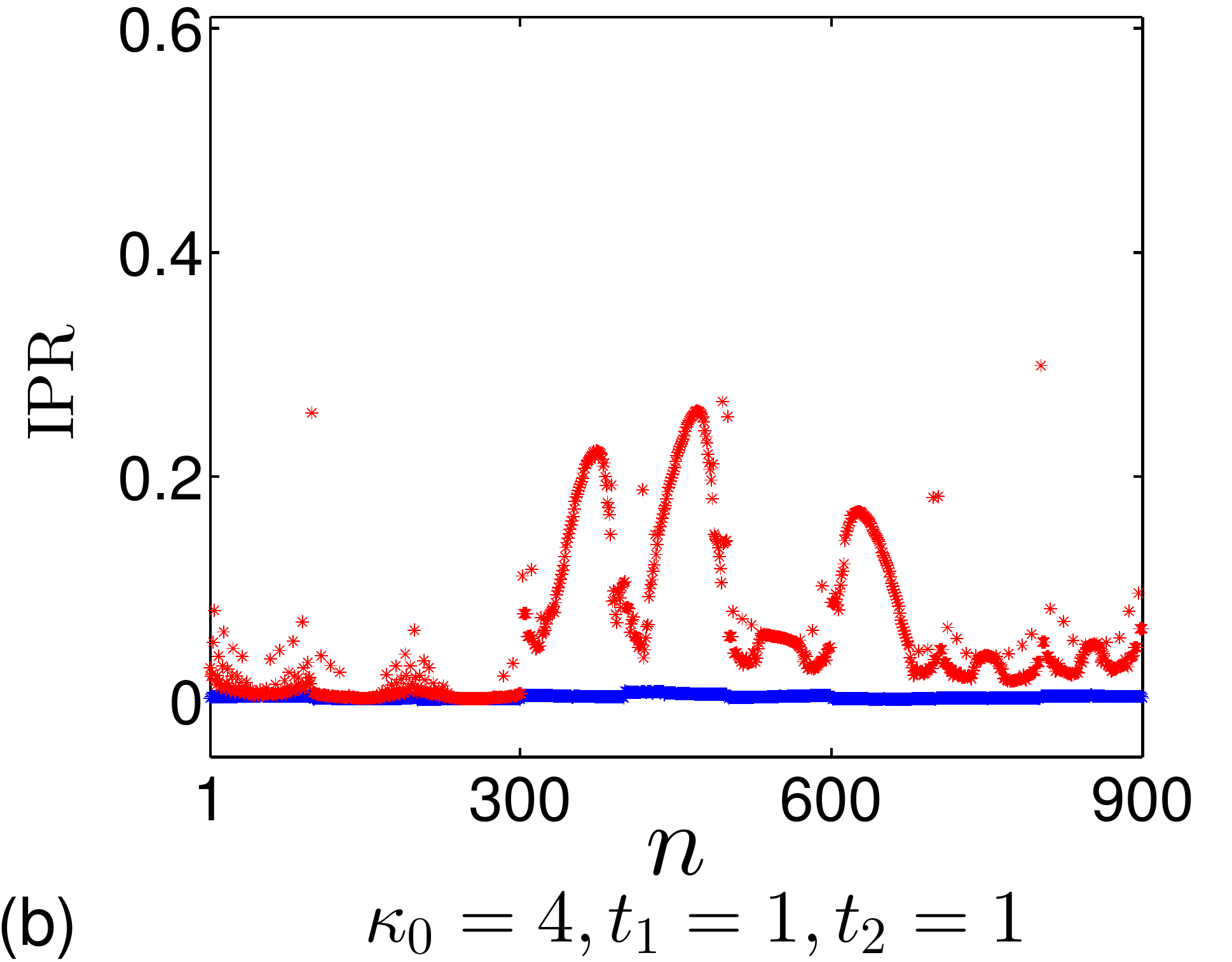}}
\end{minipage}
\begin{minipage}{0.32\linewidth}
\centerline{\includegraphics[width=5.5cm]{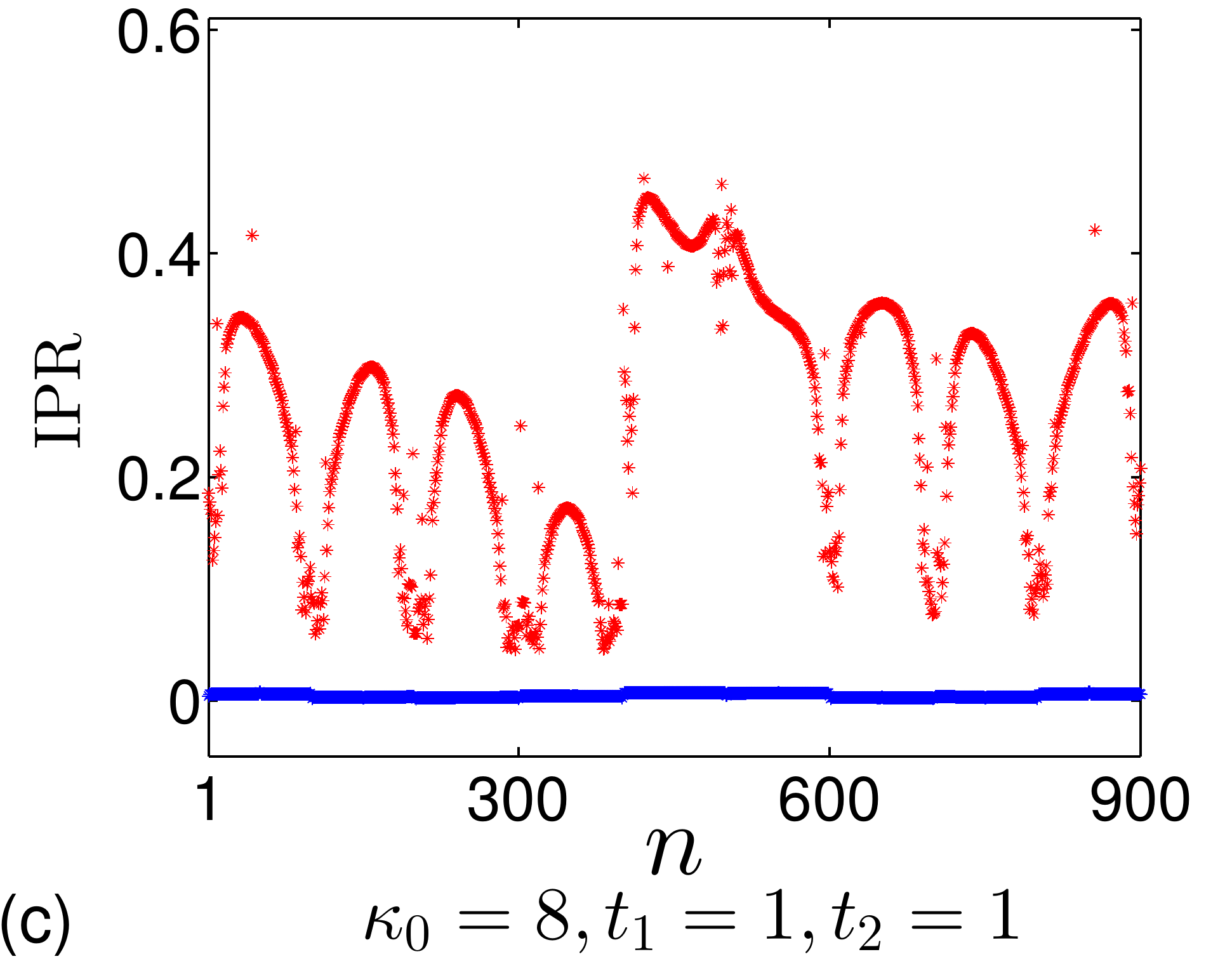}}
\end{minipage}
\begin{minipage}{0.32\linewidth}
\centerline{\includegraphics[width=5.5cm]{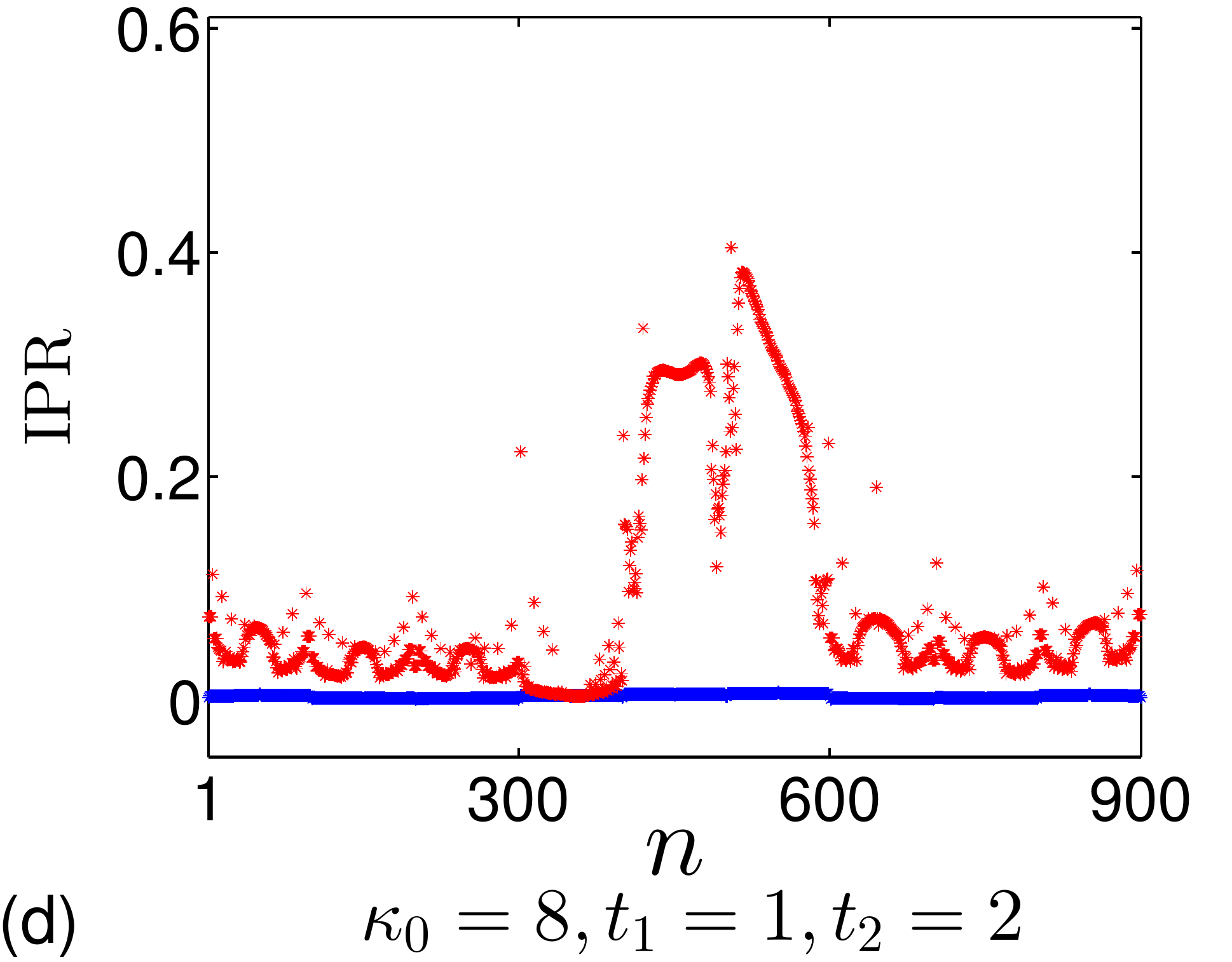}}
\end{minipage}
\begin{minipage}{0.32\linewidth}
\centerline{\includegraphics[width=5.5cm]{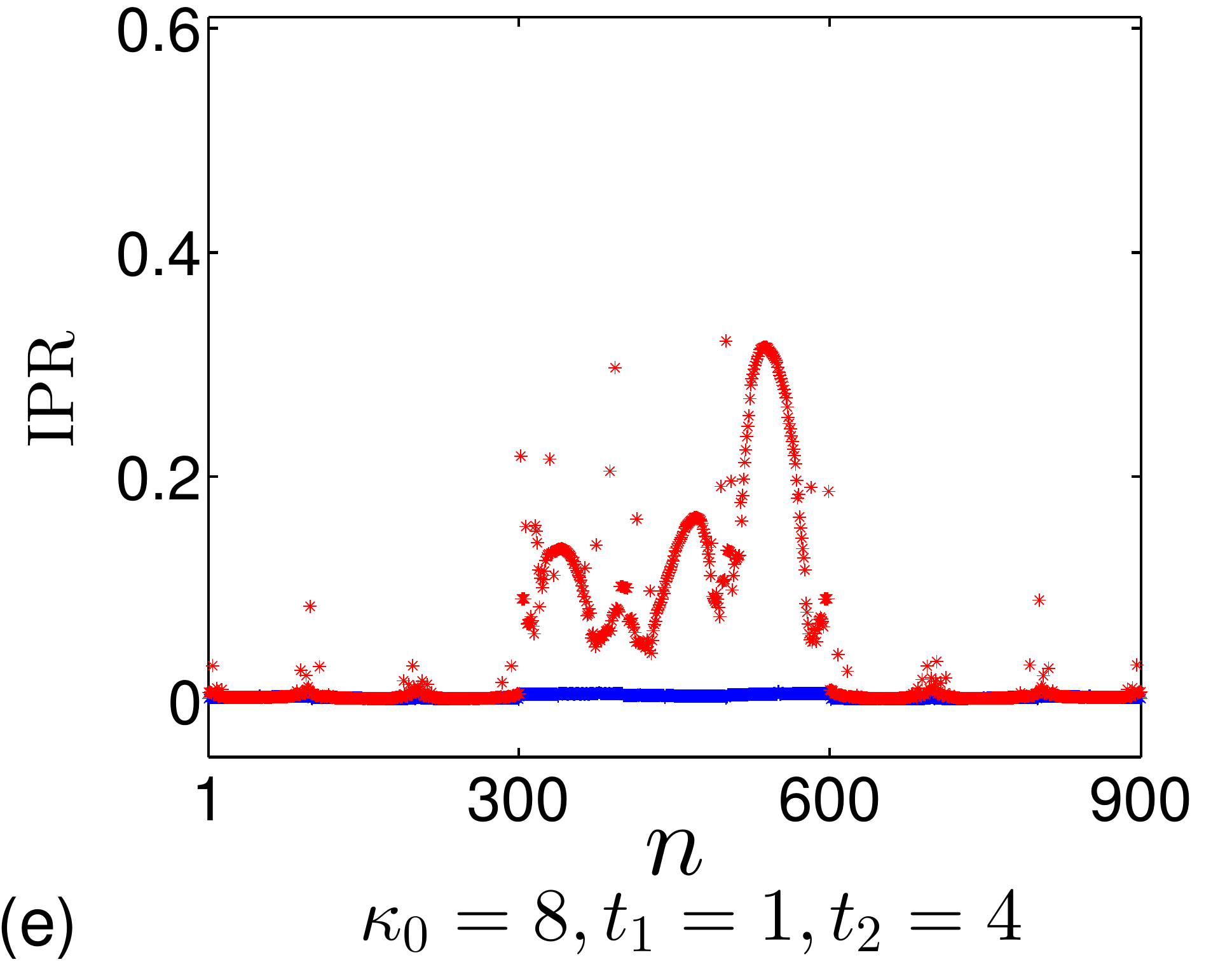}}
\end{minipage}
\begin{minipage}{0.32\linewidth}
\centerline{\includegraphics[width=5.5cm]{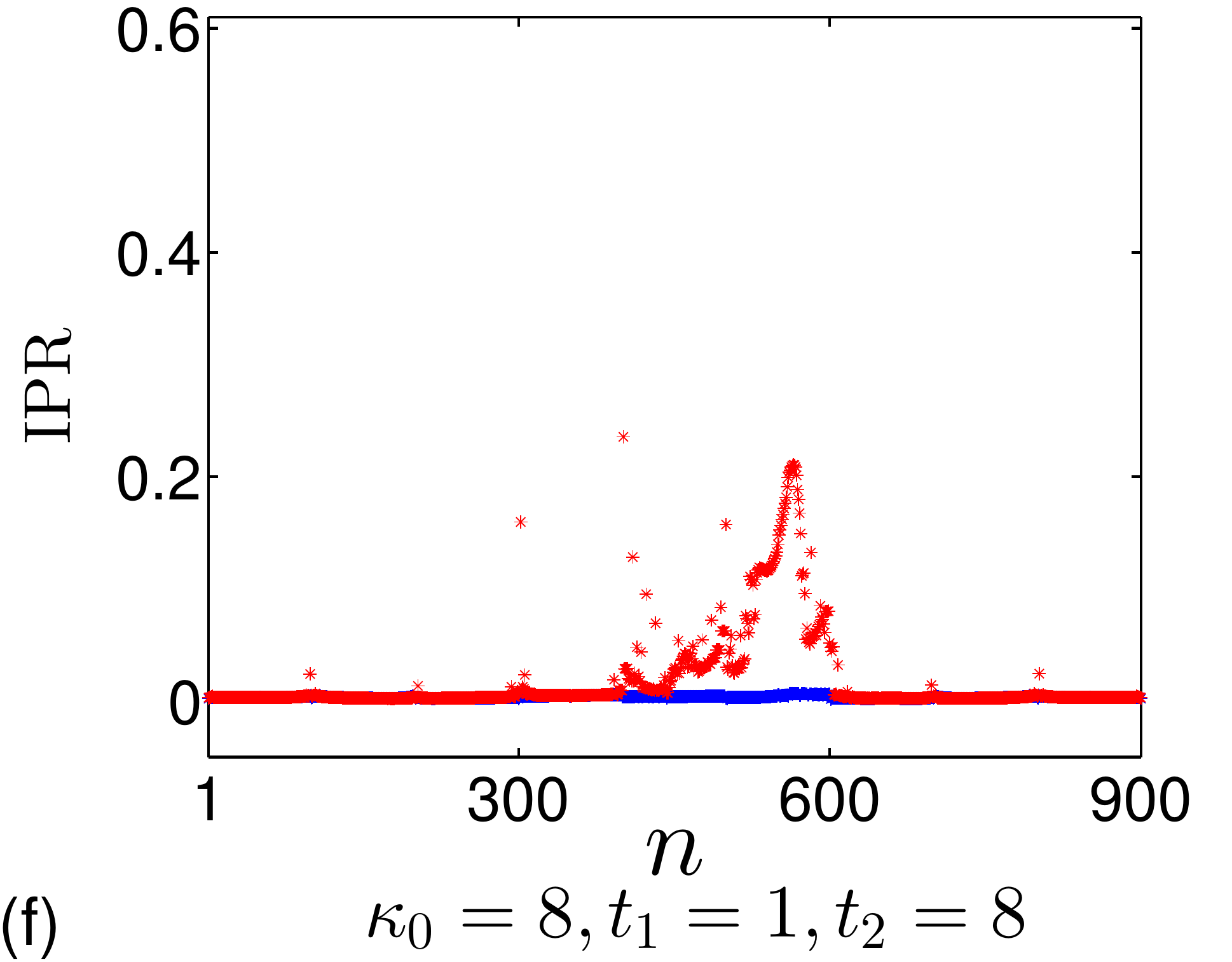}}
\end{minipage}
\begin{minipage}{0.32\linewidth}
\centerline{\includegraphics[width=5.5cm]{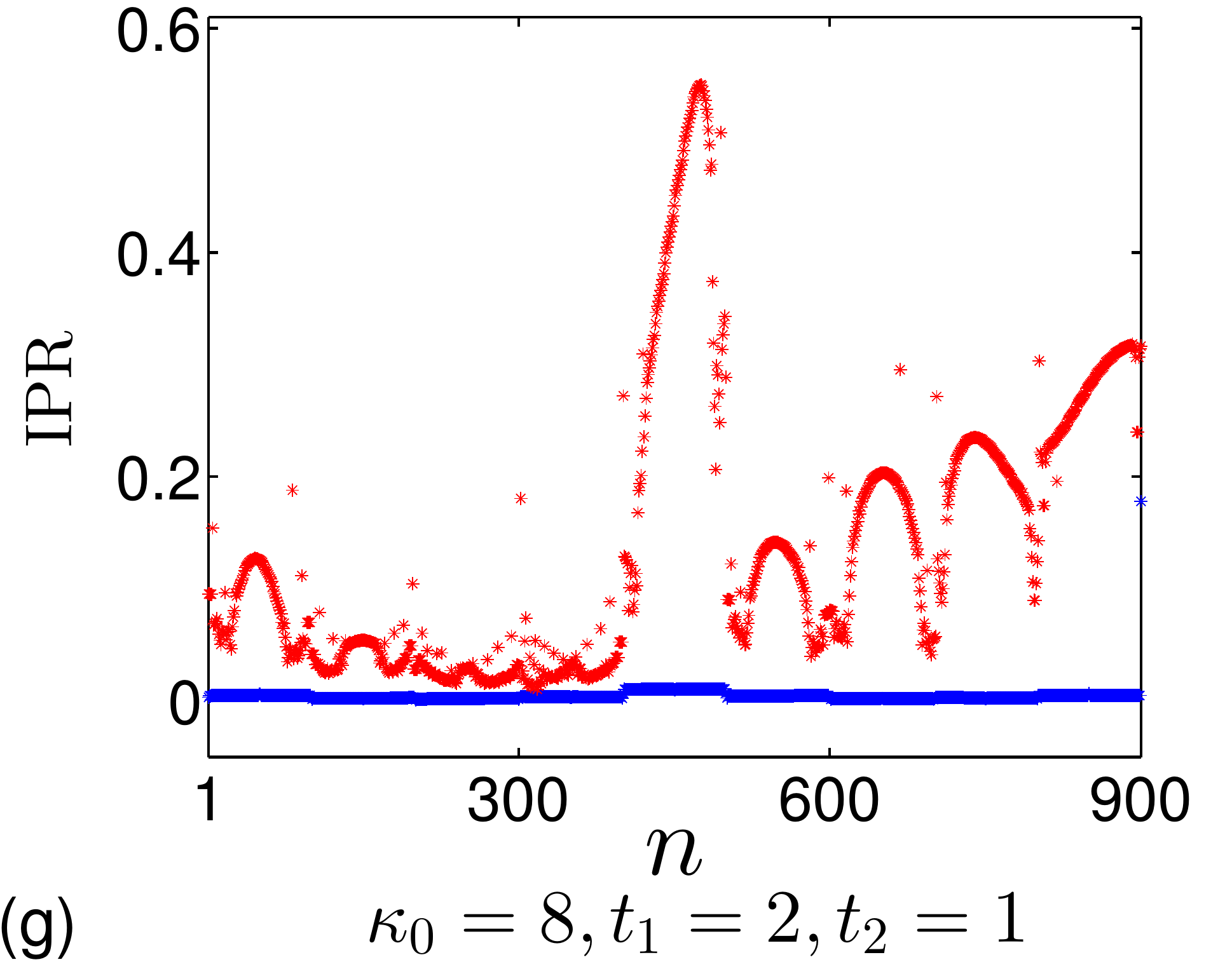}}
\end{minipage}
\begin{minipage}{0.32\linewidth}
\centerline{\includegraphics[width=5.5cm]{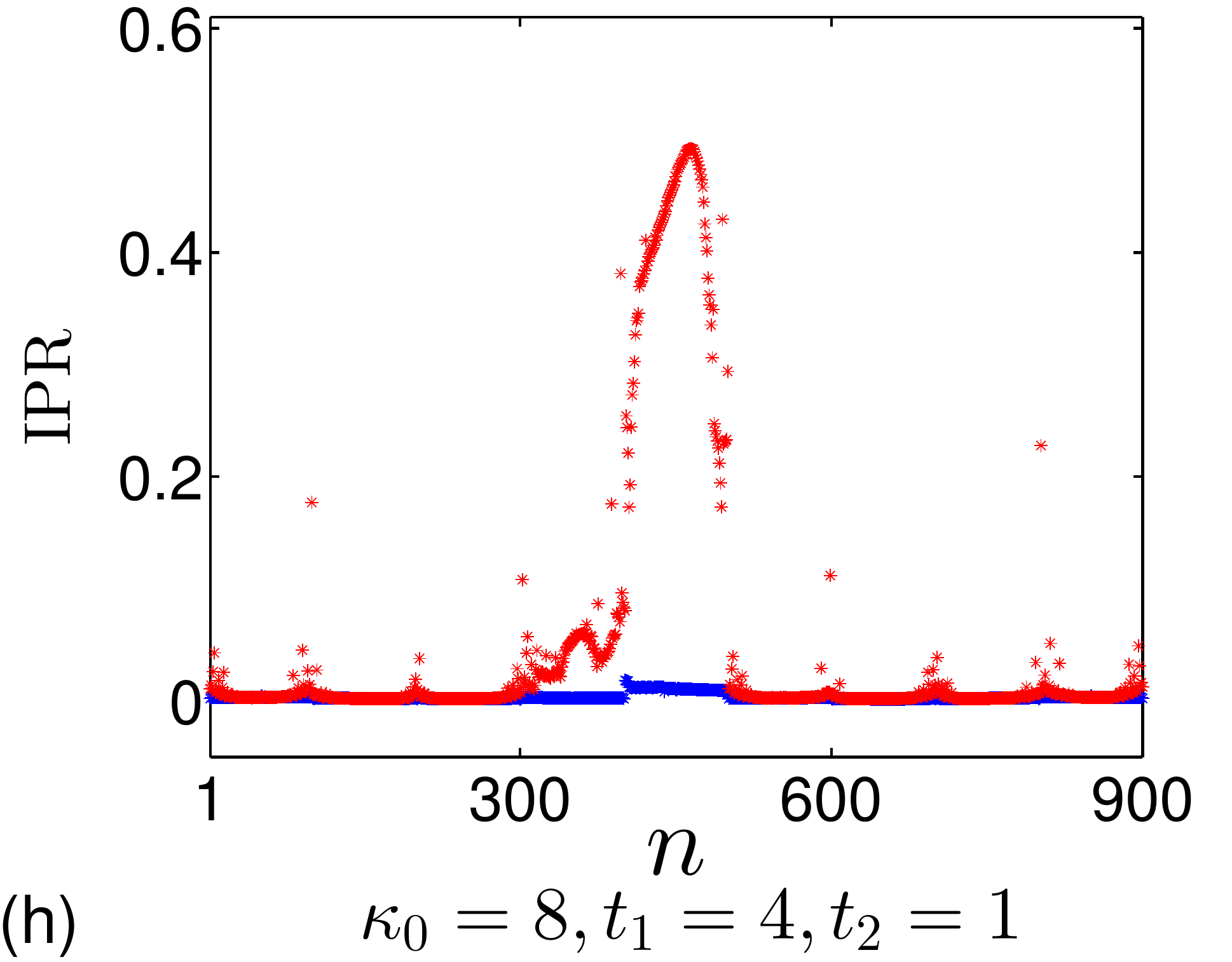}}
\end{minipage}
\begin{minipage}{0.32\linewidth}
\centerline{\includegraphics[width=5.5cm]{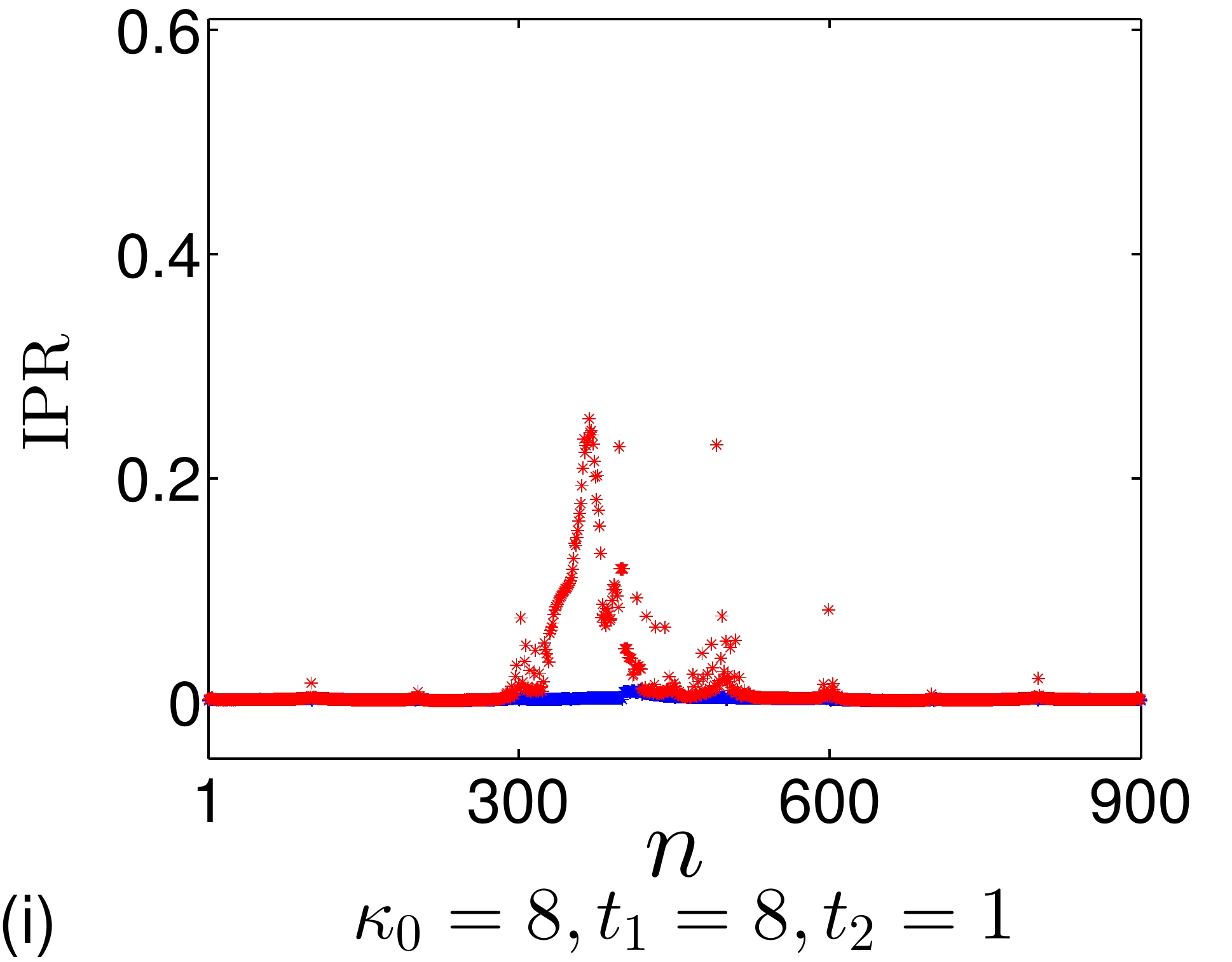}}
\end{minipage}
\caption{(Color online) Numerical simulations for the Moir\'{e} ladder
system from Eq. (\protect\ref{Moire ladd}). Plots of IPRs along energy
levels for nine typical sets of parameters both in periodic (blue star) and
quasiperiodic (red star) condition. (a), (b) and (c): $\protect\kappa _{0}$\
is $1$, $4$\ and $8$\ with $t_{1}=t_{2}=1$, respectively. (d), (e) and (f): $%
t_{2}$\ is $2$, $4$, $8$\ with $t_{1}=1$\ and $\protect\kappa _{0}=8$,
respectively. (g), (h) and (i): the same as (d), (e) and (f) but exchanging
the value of $t_{1}$\ and $t_{2}$. All \textrm{IPRs} of periodic (blue star)
ladder are trivial in this discussion. While in quasiperiodic ladder, we
note that the eigenstates of all energy levels are localized, extended or
hybrid, depending on the value of parameter $\protect\kappa _{0}$\ when the
two-legs are balanced. The mobility edge can be seen clearly in the
condition $t_{1}\neq t_{2}$\ and the performances of \textrm{IPRs} are\ also
influenced by the relative values of $t_{1}$\ and $t_{2}$. \ The size of
ladder is $N_{\mathbf{1}}=500$, $N_{\mathbf{2}}=400$, the parameter is $%
\protect\alpha =3$\ and $n$\ stands for the $n$th of energy level.}
\label{fig5}
\end{figure*}
\begin{figure*}[tbp]
\centerline{\includegraphics[width=0.8\textwidth, clip]{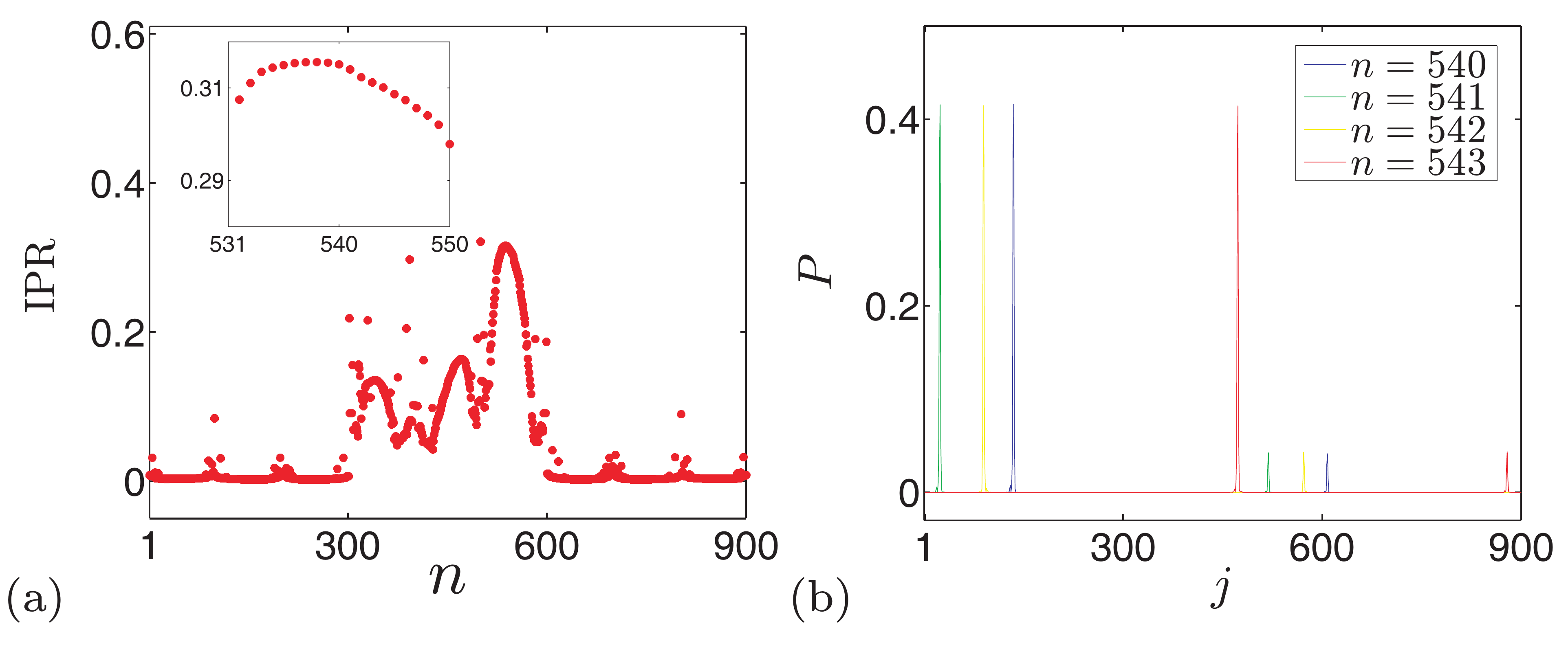}}
\caption{(Color online) (a) plot of the \textrm{IPR}s along energy levels($n$%
) for quasiperiodic ladder, as the same condition as Fig. \protect\ref{fig5}%
(e). Based on this, (b) demonstrates every two localized eigenstates labeled
by different $n$ from panel (a) have less common probability distributions. The size of
system is $N_{1}=500$ and $N_{2}=400$. The parameter $P$ is defined as $%
P=|\langle \protect\psi _{n^{\prime }}|\protect\psi _{n^{\prime }}\rangle
|^{2}$ and $j$ stands for the site of the chain.}
\label{fig6}
\end{figure*}
\begin{figure*}[tbp]
\centering
\includegraphics[width=0.75\textwidth, clip]{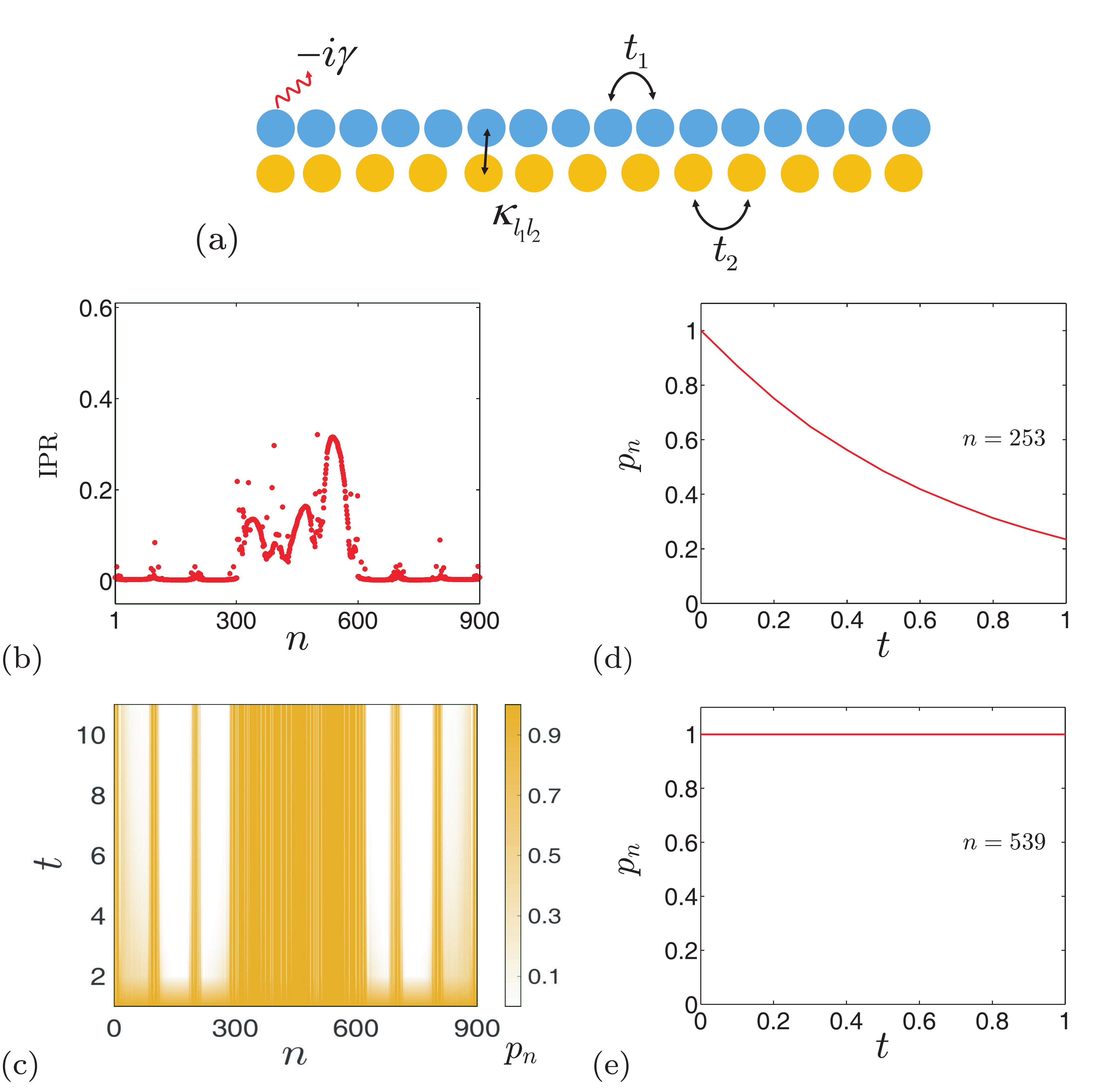}
\caption{(Color online) Schematic illustration (a) for a non-Hermitian
Hamiltonian from Eq. (\protect\ref{nH}), i.e., the non-Hermitian factor $-i%
\protect\gamma $ is induced to the Hermitian ladder. Panel (b) illustrates
the distribution of \textrm{IPR}. The corresponding mobility edge can be
detected evidently in panel (c) with the dynamical detection method in this
work. For clarity, panel (d) and (e) demonstrates the difference of
surviving probability for $|n\rangle $, which is located at extended and
localized region, respectively. Color in panel (c) stands for surviving
probability $p_{n}(t)$ and the time $t$ is in unit of $10^{3}/J$. The
parameter is $\protect\gamma =1$ and other parameters here are the same as
those in Fig. \protect\ref{fig5}(e).}
\label{fig7}
\end{figure*}

\section{Moir\'{e} ladder}

\label{Moire ladder}

In this section, we will provide a method to engineer an \textrm{AA} ladder.
In material science, moir\'{e} patterns are usually produced by stacking two
two-dimensional crystals into van der Waals heterostructures with a twist
angle. As a non-Hermitian variant of the moir\'{e} pattern, we take a simple
example by stacking two one-dimensional chains. We consider a two-leg ladder
system with the Hamiltonian%
\begin{equation}
H=H_{1}+H_{2}+H_{12}\mathrm{,}  \label{Moire ladd}
\end{equation}%
where $H_{\lambda }$\ ($\lambda =1,2$)\ describes two uniform chains%
\begin{equation}
H_{\lambda }=\sum\limits_{l}^{N_{\lambda }}t_{\lambda }\left\vert 2l,\lambda
\right\rangle \left\langle 2l+1,\lambda \right\vert +\mathrm{H.c.}
\end{equation}%
and
\begin{equation}
H_{12}=\sum\limits_{l_{1},l_{2}}\kappa _{l_{1}l_{2}}\left\vert
l_{1},1\right\rangle \left\langle l_{2},2\right\vert +\mathrm{H.c.}
\end{equation}%
represents the interchain hopping terms. Here $\kappa _{l_{1}l_{2}}$\ is the
hopping strength between $l_{1}$th site in leg $1$ and $l_{2}$th site in leg
$2$, which should be determined by the overlap of wave functions of two
particles located at two sites supposed as two (artificial) atoms in
practice, depending on the orbits of the particle and the distance between
two sites. In this paper, we simply take the Gaussian form to characterize
the corresponding interleg tunneling amplitude%
\begin{equation}
\kappa _{l_{1}l_{2}}=\kappa _{0}e^{-\alpha
^{2}[x_{l_{1}}-y_{l_{2}}]^{2}}=\kappa _{0}e^{-\alpha
^{2}[(l_{1}-1)-(1+\Delta )(l_{2}-1)]^{2}},  \label{Kappa 12}
\end{equation}%
where $x_{l_{1}}=l_{1}-1$, $y_{l_{2}}=(1+\Delta )(l_{2}-1)$\ denote the
dimensionless lattice site coordinates in chain $1$\ and $2$, respectively.
The parameter $\Delta $\ quantifies the slight difference of lattice spacing
between the two legs. The lattice spacing in leg $1$ is $1$, while $1+\Delta
$ for leg $2$. A ladder with length $L$ contains $N_{1}+N_{2}$ sites,
satisfying $L=N_{1}-1=(N_{2}-1)(1+\Delta )$. A nonzero $\Delta $\ specifies
a ladder with asymmetric legs. We note that when $\Delta $\ is a rational
number, the structure of the system is periodic and the moir\'{e} pattern
may be observed. Notably, it has quasiperiodic structure when $\Delta $\ is
an irrational number and the mobility edge may emerge. We sketch the lattice
geometries with quasiperiodic and periodic structures in Fig. \ref{fig3}.
Furthermore, the interleg tunneling strength depends on the values of $%
\Delta $\ and $\kappa _{0}$, the nonzero $\kappa _{l_{1}l_{2}}$\ is not
restricted to the nearest or next-nearest neighbour sites. As a
demonstration, we plot the distributions of hopping strengths $\kappa
_{l_{1}l_{2}}$\ in Fig. \ref{fig4} for periodic and quasiperiodic
structures, respectively.

It is expected that the performances of \textrm{IPRs} can be quite different
for the ladders with rational and irrational $\Delta $.\ To verify this
prediction, numerical simulations are given in Fig. \ref{fig5}. The profiles
of \textrm{IPRs} for irrational $\Delta $(red star) in Fig. \ref{fig5}
demonstrate the following behaviors. (i) For the case with identical
intra-leg hopping strengths $t_{1}=t_{2}$, when the interleg hopping
strength $\kappa _{0}$\ is in the order of $t_{1}$($t_{2}$), the system
possesses fully extended spectra. As $\kappa _{0}$\ increases, the mobility
edges appear, and then the spectra become fully localized. (ii) For the case
with fixed $t_{1}$ ($t_{2}$) and large $\kappa _{0}$, when $t_{2}$ ($t_{1}$)
increases, the range of localization shrinks. These results indicate that
the proposed quasiperiodic ladder system supports the mobility edge, which
is controllable by the strengths of intraleg and interleg hopping.\ It
demonstrates that the mobility edges do exist and exhibit rich structures.

\section{Dynamical detection of mobility edge}

\label{dynamical detection of mobility edge}

It emerges a natural question about how can we detect the existence\ of
mobility edge, i.e., how can we find the boundary between the localized and
extended states in experiment. We start with a dynamical method to
distinguish extended and localized spectrum region. The localized state here
has very short correlation length and every two localized eigenstates have
less common probability distributions. This feature is demonstrated by the
profiles of localized states in Fig. \ref{fig6}(b). Thus, it allows the
following dynamic behavior.\ Considering an initial state, which is the
superposition of a set of localized eigenstates.\ If these eigenstates have
no overlap in real space with each other,\ the probability distribution of
the evolved state remains unchangeably in real space. Now we turn to the
investigation for dynamical detection of mobility edge, which is defined as
the energy separating localized and extended states. We consider the time
evolution of an initial state $\left\vert \Psi \left( 0\right) \right\rangle
=\left\vert n\right\rangle $\ under a non-Hermitian Hamiltonian $H$,\ where $%
\left\vert n\right\rangle $\ is the eigenstate of the ladder system, i.e., $%
H\left\vert n\right\rangle =E_{n}\left\vert n\right\rangle $. Here $H$\
represents a modified ladder system by adding an on-site imaginary negative
potential, i.e.,%
\begin{equation}
\mathcal{H}=H-i\gamma \left\vert l,1\right\rangle \left\langle
l,1\right\vert ,  \label{nH}
\end{equation}%
where the position of the site can be arbitrarily taken. We employ the
surviving probability of $\left\vert n\right\rangle $\
\begin{equation}
p_{n}(t)=\sum\limits_{l,\lambda =1,2}^{N_{\lambda }}\left\vert \left\langle
l,\lambda \right\vert e^{-i\mathcal{H}t}\left\vert n\right\rangle
\right\vert ^{2}
\end{equation}%
to measure the mobility of the eigenstate.\ The underlying mechanism of this
method is that if the ladder $H$\ is connected by an additional infinite
system (such as an semi-infinite lead), extended state of $H$\ will escape
through the lead after sufficient long time, while a localized state of $H$\
will remain unchanged due to its uncorrelated feature (see Fig. \ref{fig6}%
(b)). An imaginary negative potential takes the role of a sink as the way
out of an extended state. It is expected that the surviving probability of $%
\left\vert n\right\rangle $\ can identify the position of mobility edge.
When $\left\vert n\right\rangle $\ is located within the extended spectrum
region, surviving probability\ decays to zero, while remains to 1 if $%
\left\vert n\right\rangle $\ within the localized spectrum region. We
perform numerical simulations to verify this scheme. We consider the
situation with the lattice geometry, parameters and plots of $p_{n}(t)$\ are
given in Fig. \ref{fig7}. In the comparison with the \textrm{IPR}
distribution, the detected mobility edges by $p_{n}(t)$\ are evident.

\section{Summary}

\label{summary}

We have shown that a quasiperiodic ladder system can support the mobility
edges, even the quasiperiodicity arises from the interleg hopping strengths,
rather than the quasiperiodic on-site potentials proposed in \textrm{AA}%
-type system. This finding opens a possibility to engineer a system
supporting mobility edge in an easy way. We have proposed a quasiperiodic
ladder system consists of two uniform chains with slight different lattice
constants (mori\'{e} ladder). An irrational lattice constant difference
results in quasiperiodic structure. Numerical simulations demonstrate that
such a system can have mobility edges. Additionally, we investigated the
dynamic detection of the mobility edge. Numerical simulations indicate that
the dynamics of the system is profoundly changed by a slight change of
lattice constant from a rational to an irrational numbers. Furthermore,
based on the measurement of surviving probability in the presence of a
single imaginary negative potential as a leakage, we have shown that the
mobility edge can be detected by a dynamic method. Our results for this
concrete system provide insightful information about the mobility edge in
quasiperiodic system. We hope that our results can stimulate experiments in
this direction.

\section*{Acknowledgement}

We acknowledge the support of NSFC (Grants No. 11874225).

\end{document}